\documentclass[runningheads]{llncs}
\usepackage{array,multirow,graphicx}
\usepackage[rgb,dvipsnames]{xcolor}

\usepackage{tikz}
\usepackage{comment}
\usepackage{amsmath,amssymb} % define this before the line numbering.
\usepackage{color}
\usepackage{hyperref}
\usepackage[accsupp]{axessibility}  % Improves PDF readability for those with disabilities.
\usepackage{orcidlink}

\usepackage[width=122mm,left=12mm,paperwidth=146mm,height=193mm,top=12mm,paperheight=217mm]{geometry}

\usepackage[normalem]{ulem}

\newcommand{\STAB}[1]{\begin{tabular}{@{}c@{}}#1\end{tabular}}

\usepackage{stackengine}
\def\tanslant{.25}
\newcommand\hatsrm[2][2]{\ensurestackMath{%
  \ifnum#1>1%
    \stackengine{-4pt}{\hatsrm[\numexpr#1-1\relax]{#2}}{%
      \scriptstyle\char'136}{O}{c}{F}{T}{S}%
  \else%
    \stackengine{-3pt}{#2}{\scriptstyle\char'136}{O}{c}{F}{T}{S}%
  \fi%
}}
\newcommand\hatsit[2][2]{\ensurestackMath{%
 \sbox0{$#2$}%
  \ifnum#1>1%
    \stackengine{-4pt}{\hatsit[\numexpr#1-1\relax]{#2}}{%
      \kern\tanslant\ht0\scriptstyle\char'136}{O}{c}{F}{T}{S}%
  \else%
    \stackengine{-3pt}{#2}{\kern\tanslant\ht0\scriptstyle\char'136}%
      {O}{c}{F}{T}{S}%
  \fi%
}}

\begin{document}
% \linenumbers
\pagestyle{headings}
\mainmatter
\def\ECCVSubNumber{6836}  % Insert your submission number here

\title{VoViT: Low Latency Graph-based Audio-Visual Voice Separation Transformer} % Replace with your title

\begin{comment}
\titlerunning{ECCV-22 submission ID \ECCVSubNumber} 
\authorrunning{ECCV-22 submission ID \ECCVSubNumber} 
\author{Anonymous ECCV submission}
\institute{Paper ID \ECCVSubNumber}

\end{comment}
%******************

% \begin{comment}
\titlerunning{VoViT: AV Voice Separation Transformer}
\author{Juan F. Montesinos\orcidlink{0000-0002-1373-3639}
\and
Venkatesh S. Kadandale\orcidlink{0000-0002-6674-0883}
\and
Gloria Haro\orcidlink{0000-0002-8194-8092}}
\authorrunning{Juan F. Montesinos et al.}
\institute{Universitat Pompeu Fabra, Carrer Roc Boronat, 138
08018 Barcelona, Spain
\email{\{juanfelipe.montesinos,venkatesh.kadandale,gloria.haro\}@upf.edu}}
% \end{comment}
%******************
\maketitle

\begin{abstract}
This paper presents an audio-visual approach for voice separation which produces state-of-the-art results at a low latency in two scenarios: speech and singing voice. The model is based on a two-stage network. Motion cues are obtained with a lightweight graph convolutional network that processes face landmarks. Then, both audio and motion features are fed to an audio-visual transformer which produces a fairly good estimation of the isolated target source. In a second stage, the predominant voice is enhanced with an audio-only network.
%\gh{The main contributions are i) Presenting a low-latency SOTA AV voice separation model, ii) showing that refinement networks switching the optimisation problem to solve can work better than training larger models, iii) showing that face landmarks can compete against raw video and help to deal with large-scale datasets and iv) a set of ablation studies and comparison to state-of-the-art methods in which we explore the transferability of  models trained for speech separation in the task of  singing voice separation.}
{We present different ablation studies  and comparison to state-of-the-art methods. Finally, we explore the transferability of  models trained for speech separation in the task of  singing voice separation.}
The demos, code, and weights {are available in \url{https://ipcv.github.io/VoViT/}}. %\url{https://av-voice-sep.github.io}  
\end{abstract}

\section{Introduction}
%Multimedia content has grown exponentially in the last decades. Social networks such as Youtube, Instagram or Tiktok have lead to millions of hours of unprocessed data. Processing these kind of data involves several tasks such as scene understanding, summarization, classification, retrieval, speech recognition and enhancement, among others. Many approaches are based on the joint processing of both audio and visual modalities. In this paper we address the voice separation and enhancement problems from a multimodal perspective, leveraging the visual stream to guide the resolution of the problem. \gh{}{we can remove this paragraph and safe space}

Human voice is usually found together with other  sounds. Think of people speaking in a cafeteria or in a social gathering, a journalist reporting on the scene, or an artist singing on a  stage. In these situations we can find: multiple concurrent speeches, speech with background noise or a single or multiple singing voices with music accompaniment among others. Our brain is capable of understanding and concentrating on the voice of interest \cite{cherry1953some}. This cognitive process does not only rely on the hearing. Some works have shown the sight helps to focus on the voice of interest  \cite{ZionGolumbic2013VisualIE} or to resolve ambiguities in a noisy environment \cite{ma2009lip}.  In this paper we address the voice separation and enhancement problems from a multimodal perspective, leveraging the motion information extracted from the visual stream to guide the resolution of the problem.

%\gh{Most of the works on speech separation (see Section \ref{related}) show a consistent improvement by using audio-visual methods over audio-only ones. However, there is no methodical analysis on the extra computational cost of those audio-visual methods vs the performance obtained. *to check*}{}

We propose an audio-visual (AV) voice separation model that produces state-of-the-art results. It is based on a two-stage approach. The first stage estimates a fairly good separation by combining audio and motion features with a transformer. Motion cues are crucial when the sound mixture contains different predominant voices. We extract those cues with a graph convolutional network (CNN) that processes a sequence of face landmarks. The audio-visual features are aligned in the feature dimension and preserve the time resolution. They are processed by a multimodal spectro-temporal transformer that estimates the isolated voice corresponding to the target face landmarks. In a second stage, the predominant voice is enhanced by a small audio-only U-Net that takes as input just the pre-estimated audio. The voice of interest is predominant in the first estimation and thus an audio-only network is capable of modelling it and cancelling the sparse and mild interferences present in the pre-estimation. The paper includes an ablation study of different configurations of the multimodal transformer, its number of blocks and design of the lead voice enhancer network. 
The proposed method is compared to state-of-the-art methods in two different scenarios: speech and singing voice separation, showing successful results in both cases.

The contributions of this work are several: i) We propose an audio-visual network based on a transformer which performs better than current state-of-the-art models in speech and singing voice separation.  ii) We show that a landmark-based approach for extracting motion information can be a lightweight competitive alternative to processing raw video frames. iii) We show how an enhancement stage based on a light network can boost the performance of AV models over larger complex models, reducing the computational cost and the required time for training. iv) We reveal that AV models trained in speech separation do not generalise good enough for the separation of singing voice because of the different voice characteristics in each case and that a dedicated training with singing voice examples clearly boosts the results.  Finally, v) our method is an  end-to-end gpu-powered system which is capable of isolating a target voice in real time (including the pre-processing steps). 
%We study the performance vs speed of several sound source separation methods and b) We propose a series of lightweight AV models which are SOTA in voice separation. 

\section{Related work} \label{related}
In the last years there has been a fast evolution of deep-learning-based  audio-visual works for speech separation and enhancement (we refer the reader  to a recent review in \cite{michelsanti2021overview}).

Back in 2016, we can find one of the first works in exploiting visual features for speech enhancement \cite{wu2016multi}. In this work, the authors proposed a CNN to process the visual signal and a fully connected layer to process the raw waveforms. Both modalities were fused by a BiLSTM network. This network had approximately 3M parameters (M for millions), far from the 80M of the most recent work \cite{visualvoice}. A two-tower stream for processing audio and video features and then fused with a BiLSTM module that predicted complex masks was proposed in  \cite{ephrat2018looking}.

A two-step enhancement process was proposed in \cite{Afouras18}. In the first step, a two-tower stream processed the audio-visual information to extract a binary mask that performed separation on the magnitude spectrogram. Afterwards, the  phase of the spectrogram was predicted by passing the estimated magnitude spectrogram together with the noisy phase spectrogam through a 1D-CNN. A similar idea was developed in  \cite{gabbay2018visual}, where a two-tower stream encoder generated  an embedding of audio-visual features from which the enhanced speech spectrogram was recovered. On the other hand, in \cite{hou2018audio} not only the enhanced spectrogram was reconstructed but the input frames as well.

New approaches and explorations different from the two-tower CNNs appeared recently.  Variational auto-encoders \cite{sadeghi2021mixture} for  speech enhancement joined the scene. Concurrently, \cite{wu2019time} developed a time-domain model for speech separation, in contrast to most of the works which usually posed the problem in the time-frequency domain. 
 Multi-channel audio-visual speech separation was addressed in \cite{gu2020multi} in a four-tower stream fashion. The mixture spectrogram was constrained with directional features from the visual stream of the speaker. A temporal CNN extracted visual features from the lips motion.  The audio and visual embeddings were concatenated together with a speaker embedding extracted from the clean audio(s). 
 A different mechanism was used in \cite{9054180,sato2021multimodal,sun2020attention}, where the audio-visual fusion was done with an attention module; or in \cite{xu2021vsegan}, where the system was trained in a GAN manner so that the discriminator modeled the distribution of the clean speech signals. 
 Transformers have been used in audio-only source separation \cite{spectral_transformer}. 
 %, instrument sep, singing voice sep).}\textcolor{red}{COMPLETE} 
 Very recently, audio-visual transformers  were investigated  in \cite{truong2021right} for main speaker localization and  separation of its corresponding audio.
In \cite{tzinis2021improving} an audio-visual transformer was used for classification in order to guide an unsupervised source separation model. Finally, in  \cite{chen2021audio} a transformer was used for audio-visual synchronisation. 
 
 Another interesting proposal is \cite{chuang2020lite}, where the authors were concerned about the extra computational cost of processing the visual features and the possible privacy problems arised from it. 
 On the other hand, to our knowledge, there are only two works using face landmarks, instead of video frames, for source separation. In \cite{morrone2019face} they process face landmarks with fully connected layers and then use BiLSTMs to predict the masks for the target source. %and explored different ways of leveraging the visual features.
 In \cite{montesinos} a U-Net conditioned by a graph convolutional network that processed face landmarks was used for audio-visual singing voice separation.
The work in \cite{michelsanti2019training} compared different training targets and loss functions for audio-visual speech enhancement.

Most recent algorithms made use of lips motion as well as appearance information, usually implementing cross-modal losses to pull together corresponding audio-visual features \cite{visualvoice,makishima2021audio}.

\section{Approach}
In audio-visual voice separation, given an audio-visual recording with several speaking/singing faces, 
{and other sound sources,} 
%with or without any other accompaniment, 
the goal is to recover their isolated voices by guiding the voice separation with the visual information present in the video frames. 
%More formally, given a set of speakers $J=(V,S)$ where $V=\{v_i(t) \, | \, i=1, ..., N\}$ is the visual signal and $s=\{s_i(t) \, | \, i=1, ..., N\}$ is the voice signals, 
More formally, given the audio signal of each speaker, $s_i(t)$ (where $t$ denotes time), 
the mixture of sounds can be defined as $x(t)=\sum_i s_i(t)+n(t)$ where $n(t)$ denotes any other sound present in the mixture, i.e. background sounds. Therefore, the task of interest can be defined as the estimation of each individual voice $\hat{s_i}(t)$. In our approach $\hat{s_i}(t)=F(x(t),v_i(t))$, where $F$ is a function represented by a neural network. The network receives the visual information of the speaker of interest, $v_i(t)$, and estimates its isolated voice $\hat{s_i}(t)$.

\subsection{The AV Voice Separation Network}
\begin{figure}[ht]
  \centering
  \includegraphics[width=\textwidth]{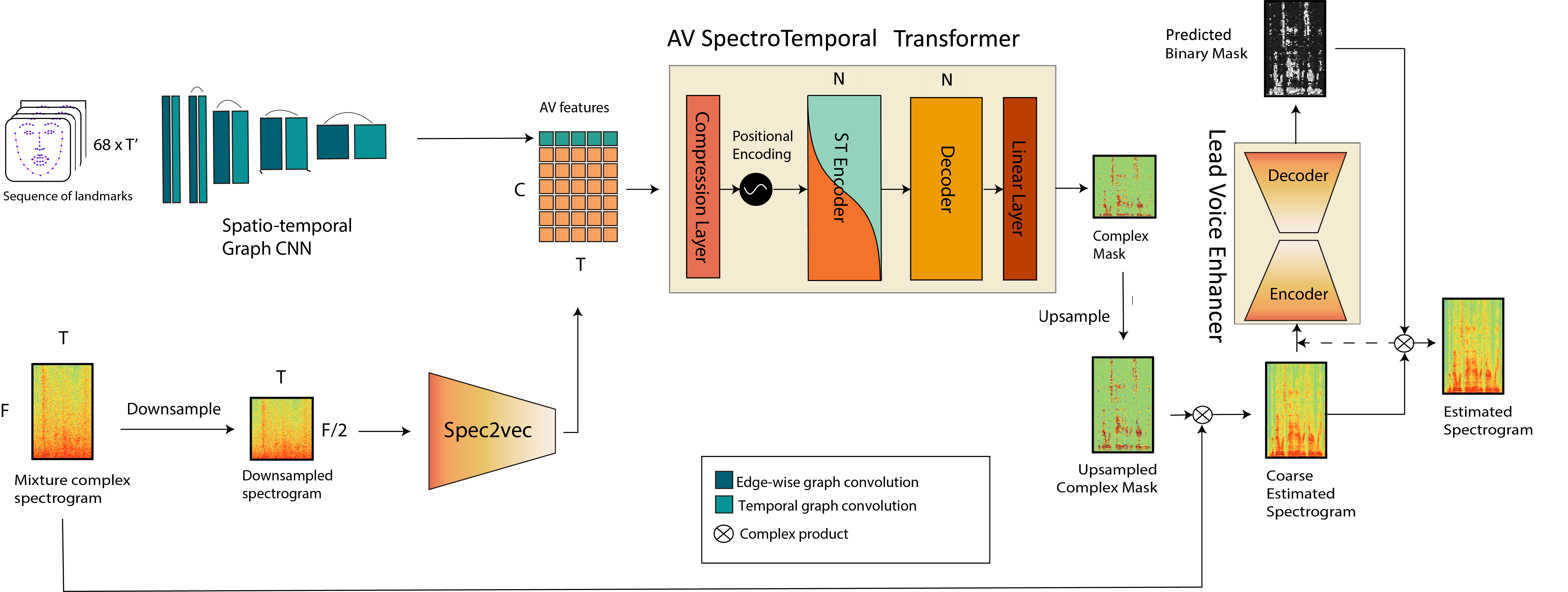}
  \caption{Audio-visual voice separation network. Audio and video features are concatenated in the channel dimension before being fed to the transformer.}
  \label{img_model}
\end{figure}

Our solution comprises of a two-stage neural network that operates in the time-frequency domain. The first stage consists of an AV voice separation network which can isolate the target voice at a good quality. However, this network is the most demanding one in terms of computational cost. To alleviate this, we propose to use downsampled spectrograms in this stage. The second stage consists of a recursive lead voice enhancer network that works with full resolution spectrograms.  In Section  \ref{sec:refinement}, we experimentally show that this two-stage design leads to a higher performance than using larger AV models. %\textcolor{red}{ Besides, the user has control on the latency/performance trade-off by disabling the refinement stage or reducing its depth.} 
To achieve this modularity, the networks at both stages are trained independently.  The whole model is presented in Fig.\ref{img_model}.  

{\bf Stage 1: Audio-Visual Voice Separation.} \label{stage-1}
For simplicity, we seek to isolate the voice (denoted by $s(t)$ and its corresponding spectrogram $S(f,t)$) corresponding to a single face at a time.
The audio waveform of the mixture, $x(t)$, is transformed into a complex spectrogram $X(f,t)$ applying the Short-Time Fourier Transform (STFT). 
Once the waveform is mapped to the time-frequency domain, we can define a complex mask $M(f,t)$ that allows to recover the spectrogram of the estimated source with a complex product, denoted as $*$, that is: 
$S(f,t)=X(f,t)*M(f,t)$
Then, the goal of the network in the first stage  %$f_\theta$ 
is to estimate the complex mask $\hat{M}(f,t)$. %$M(f,t)$, that is $\hat{M}=f_\theta(x(t),v_i(t))$. \gh{}{(not sure if use $x$ and $v_i$ or the spectrogram $X$ and the seq. of graphs)}
The optimal set of parameters of the network  %$\theta$ 
is found by minimising the following loss:
%\begin{equation*}
%       \theta= \arg  \underset{\theta}{\min}||G\odot (\mathcal{X}_b-q(\mathcal{V},\mathcal{M}))||^2_2
%\end{equation*}
\begin{equation*}
       \mathcal{L}_1 =  \| G\odot (M_b-\hat{M}_b) \|^2
\end{equation*}
where $M_b$ and $\hat{M}_b$ are, respectively, the ground truth and estimated bounded complex masks,
%\gh{}{*I think $\mathcal{V}$ has not been defined, same for the other variables. $q$ is both the network and parameters .... I would say $f(X; \theta)$ for the network and $\theta$ for the parameters, as it is common in different papers, then we can talk about the optimal set of parameters $\theta^*$. Another think: why here $\mathcal{M}$ and then $M$? I find the notation. and lack of definition of variables very confusing*}
$\odot$ denotes the element-wise product, $\|\cdot\|$ is the $L2$-norm and $G$ is a gradient penalty term which weights the time-frequency points of the mask according to the energy of the analogous point in the mixture spectrogram $X$:
\begin{equation}\label{eq:G}
    G(f,t;X) = \max(\min(\log(1+\| X(f,t) \|),10),10^{-3}).
\end{equation}
Note that, by definition, the ground truth mask $M$ is not bounded. In order to stabilise the training, we bound the complex masks by applying a hyperbolic tangent \cite{williamson2015complex}:
$M_b = \tanh{M^r}+i \tanh{M^i}$, where $M^r$ and $M^i$, denote the real and imaginary parts, respectively. The audio waveform of the estimated source can be computed through the inverse STFT of the estimated spectrogram $\hat{S}(f,t)=X(f,t)*\hat{M}(f,t)$.

To solve the AV voice separation problem, we propose to leverage the face motion information present in the video frames of the target person whose voice we want to isolate. For that, we use a spatio-temporal graph neural network that processes the face landmarks to generate motion features. On the other hand, the audio features are generated by a CNN encoder, denoted as \textit{Spec2vec}. Both audio and motion features preserve the temporal resolution and are concatenated in the channel dimension, then they are fed into a transformer. All the submodules have been carefully designed to achieve a  high-performance low-latency neural network. 

\textbf{Spatio-temporal graph CNN:} Many AV speech separation or enhancement methods  rely on lips motion extracted from raw video frames to guide the task.  To reduce the computational cost of the visual stream, we propose to use face landmarks together with a spatio-temporal graph CNN \cite{yan2018spatial}. This network, similar to that in \cite{montesinos}, was redesigned to preserve the temporal resolution. It consists of a set of blocks which apply a graph convolution over the spatial dimension followed by a temporal convolution. This way we can considerably reduce  the amount of data to process and to store, from  $96\times 96\times 3\approx 3\cdot10^4$ values per frame to $68\times 2\approx 10^2$. This supposes a substantial reduction in the storage necessities when working with large audio-visual datasets. For example,  \textit{Voxceleb2}'s  grayscale ROIs occupy 1Tb, the raw uncompressed dataset occupies several Tb while storing face landmarks only requires 70 Gb.

\textbf{Spec2vec:} It is well known that transformers need proper embeddings to achieve high performance. We use the audio encoder of \cite{ephrat2018looking} to generate embeddings without losing temporal resolution.   

\textbf{AV spectro-temporal transformer: }  The traditional AV source separation methods comprise of a two-tower stream architecture. We can find two major variants: either encoder-decoder CNNs (usually with a U-Net as backbone) (e.g. \cite{visualvoice,gao2019co,montesinos,slizovskaia2021conditioned,zhao2018sound,zhao2019sound}) or recurrent neural networks (RNNs), both conditioned on visual features (e.g. \cite{ephrat2018looking,morrone2019face,wu2016multi}). The major drawback of the latter is that RNNs are sequential, introducing bottlenecks in the processing pipeline. Transformers appeared as an efficient solution, reaching the same performance than RNNs and CNNs in large datasets. They are trained with a masking system allowing to process all the timesteps of a sequence in parallel. However, these architectures operate sequentially at the time of inference, like the RNNs. To overcome this issue we use an encoder-decoder transformer, which can solve the source separation problem in a single forward pass.

Transformers were originally designed to work with two unimodal signals. We study three different possible configurations for the  transformer. 
The first proposal is to use the transformer as an auto-encoder, being fed with an audio-visual signal directly. This way we ease the task for the transformer as audio and visual features are temporally aligned by construction. Then, it just has to find relationships through the multi-head self-attention. 
The second proposal is to pass visual features to the encoder and audio features (from the mixture) to the decoder so that the network can find audio-visual interdependencies via multi-head attention. Nevertheless, we hypothesise the dependencies between video and audio are local as audio events mostly occur at the same time than visual events. 
Lastly, we feed the encoder with an audio-visual signal and the decoder with the ground-truth separated audio. Note that this model is slower than previous ones as the model runs recurrently at inference time, going from a time complexity of $\mathcal{O}(n)$ to $\mathcal{O}(n^2)$ where $n$ is the  length of the sequence.
From the ablation study in Section \ref{multimodal-experiment} and Table \ref{transfomer-ablation}, we conclude that the best model is the first one, i.e. the one that uses an audio-visual signal as input, we denote it as AV ST-transformer.

\begin{figure}[ht]
  \centering
  \includegraphics[width=0.8\textwidth]{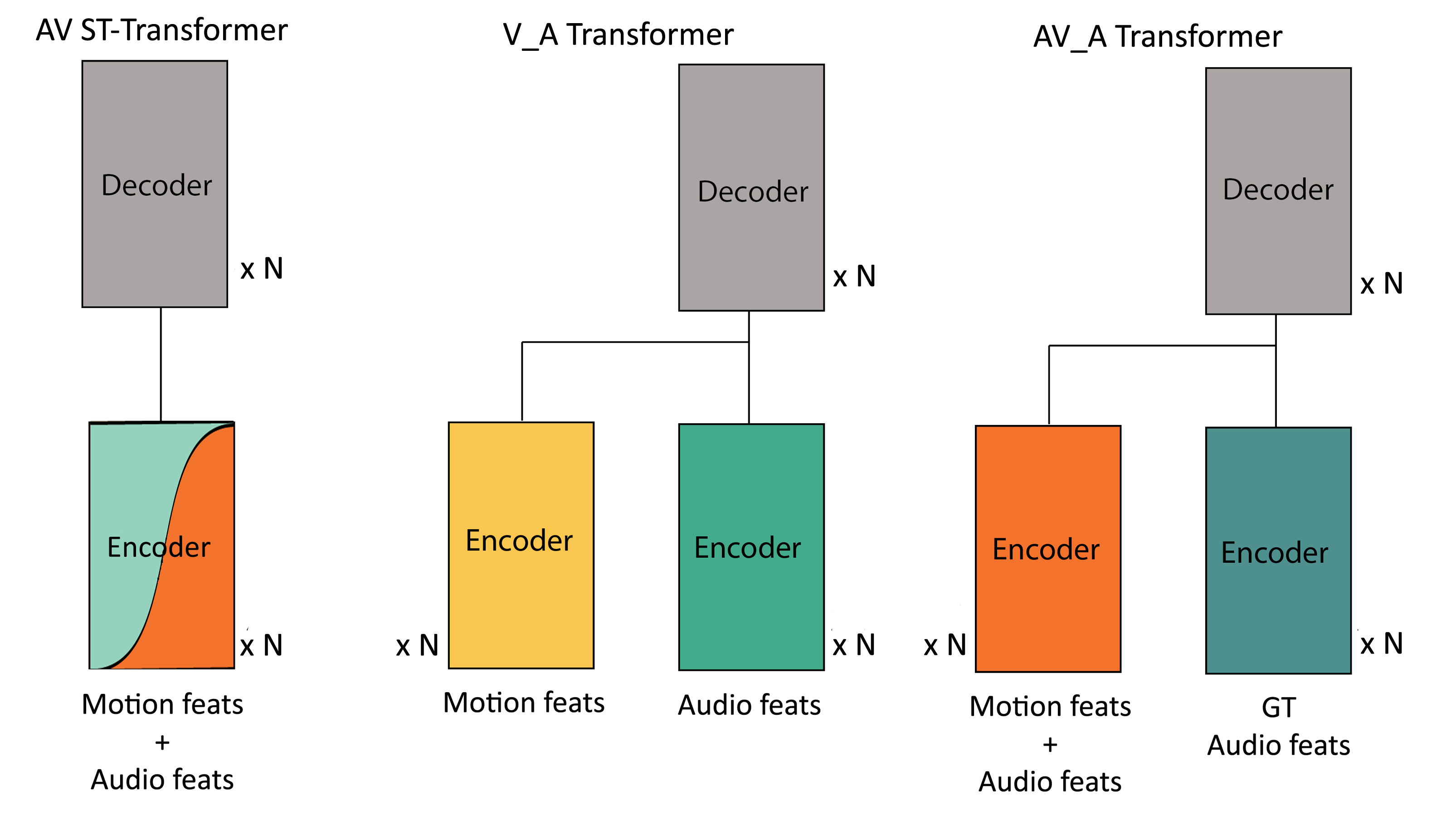}
  \caption{Three proposed ways to feed a transformer with an audio-visual signal. Left: audio-visual signal, middle: video to the encoder and audio mixture to the decoder, right: audio-visual signal to the encoder and clean audio to the decoder. }
  \label{transformer-ablation}
\end{figure}

We design our AV ST-transformer encoder upon the findings of \cite{spectral_transformer}. The AV ST-transformer has 512 model features across 8 heads. We tried 256 features but it works worse. The compression layer is nothing but a fully connected layer followed by GELU \cite{gelu} activation which maps the $C$ incoming channels to the 512 channels required by the architecture. It is composed by $M$ encoders and $M$ decoders. The encoder is a set of two traditional encoders in parallel, which processes the signal  from a temporal and a spectral point of view \cite{spectral_transformer}.  %The transformer is illustrated in the Fig. \ref{fig:spec-trans}.
%\begin{figure}[ht]
%  \centering
%  \includegraphics[width=0.8\textwidth]{imgs/spectrans_rot.png}
%  \caption{Design of the VoViT's transformer encoder block. Illustration from \cite{spectral_transformer}.}
%  \label{fig:spec-trans}
%\end{figure}

{\bf Stage 2: Lead voice enhancer.}  \label{approach:refinement}
%One of the main flaws of audio-visual models arises from the ambiguities between lips motion and their corresponding phonemes. This implies several phonemes are compatible to a given viseme and the models may not to filter those.
Although lips motion is correlated with the voice signal and may help in source separation,  it is not always accessible or reliable. For example, the scenarios involving a side view of the speaker or a partial occlusion of the face or an out-of-sync audio-visual pair make it challenging to incorporate the lips motion information in a useful way; all such scenarios may appear in unconstrained video recordings. 
In \cite{montesinos}, the authors show that audio-only models tend to predict the predominant voice in a mixture when there is no prior information about the target speaker. Based on this idea, we hypothesise that, if the first stage of the AV voice separation network outputs a reasonable estimation of the target voice, this voice will be predominant in the estimation. Upon this idea, we use an audio-only network which identifies the predominant voice and enhances the estimation without relying on the motion, just on the pre-estimated audio. To do so, we simply use a small U-Net which takes as input the estimated magnitude spectrogram (at its original resolution) and returns a binary mask. The ground truth binary mask can be obtained from the ground truth spectrogram $S$ and the spectogram to be refined, $\hat{S}$, which is the one  estimated in the stage 1:

%\fix{the pre-estimated spectrogram $\hat{S}$}{$\hat{S}$, which is the spectrogram estimated in the stage 1,} and the ground truth spectrogram $S$ as follows
\begin{equation}\label{bmask}
        M(f,t) =  \begin{cases}
1, & \text{if} \   \|S(f,t)\|\ge \|\hat{S}(f,t)-S(f,t)\|, \\
0, &   \text{otherwise}.
\end{cases}
\end{equation}
Notice that the difference $\hat{S}(f,t)-S(f,t)$ are the remaining sources that need to be removed in the refinement stage.

There are different reasons to use binary masks. On the one hand, we found qualitatively, by inspecting the results, that the secondary speaker is often attenuated but not completely removed.  In \cite{grais2016combining}, the authors show that binary masks are particularly good at reducing interferences.
On the other hand, complex masks appeared as an evolution of binary masks and ratio masks, as a way of estimating, not only the magnitude spectrogram, but the phase too. Note that these masking systems usually reconstruct the estimated waveform with the phase of the mixture as they estimate the magnitude only. In our case, the phase has already been estimated by using complex masks in the previous stage. Lastly, by using binary masks, we are changing the optimisation problem and easing the task since it is simpler to take a binary decision than orienting and modulating a vector.

Note that this refinement network can run recursively, although we empirically found (see Table \ref{tab:refinement}) that applying the refinement network once leads to the best results in terms of SDR and a considerable boost in SIR. Further iterations reduce the interferences (at a lesser extent) but at the cost of introducing more distortion.

Let us denote by $\hatsit[2]{M}$ the binary mask estimated by the lead voice enhancer  network. We trained this network to optimise a weighted binary cross entropy loss:
\begin{equation*}
       \mathcal{L}_2=\!\sum^F_{f=1}\sum^T_{t=1} \frac{G(f,t;\hat{S})}{FT} \left( M(f,t)\log{\|\hatsit[2]{M}(f,t)\|+(1-M(f,t))(1-\log{\|\hatsit[2]{M}(f,t)\|)}}\right)
\end{equation*}
where the weights $G$ are defined in \eqref{eq:G}.
%\begin{equation*}\label{G}
 %   G(f,t) = \max(\min(\log(1+||\hat{S}(f,t)||),10),10^{-3})
%\end{equation*}

\subsection{Low-latency data pre-processing}
Many audio-visual works rely on expensive pipelines to pre-process data, which makes the proposed systems unusable in a real-world scenario unless a great amount of time is invested in optimisation. Pursuing the real applicability of our model, we curated an end-to-end gpu-powered system which can  pre-process (from raw audio and video) and isolate the target voice of 10s of recordings in less than 100ms using floating-point 32  precision, and in less than 50 ms using floating-point 16 precision.

\begin{figure}[ht]
  \centering
  \includegraphics[width=0.7\textwidth]{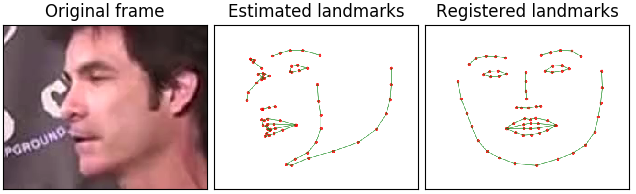}
  \caption{Frame example from \textit{Voxceleb2} \cite{voxceleb} with partial occlusions. Thanks to the landmark estimation together with the registration we can estimate the unoccluded lips. }
  %$Trace= id06816/6pDTN5v_wfE/00014 frame 0$
  \label{side-view}
\end{figure}

\textbf{Face landmarks: }The most common approach in speech separation is to {align the faces in the different frames} via 2D face landmark estimation together with image warping (e.g. \cite{visualvoice,montesinos}). This step removes eventual head motions.
In order to achieve real-time audio-visual source separation, we estimate the 3D face landmarks using an optimised version of \cite{faceal} and
%\fix{register the face face landmarks to a frontal view omitting the image warping}{obtain \gh{a}{ 
an aligned  frontal view by applying a rigid transformation, skipping the image warping step. This optimised preprocessing takes around 10 ms to process 10s of video. 
Thanks to the 3D information, we can recover lips motion from side views  by estimating 3D landmarks, as shown in  Fig. \ref{side-view}. To do the registration, we use the Kabsch algorithm \cite{Kabsch}. Finally, we drop the depth coordinate and consider just the first two spatial coordinates in the nodes of the graph. 
% A Procrustes transformation is a geometric transformation that involves only translation, rotation, uniform scaling, or a combination of these transformations.

\textbf{Audio:} Waveforms are re-sampled to 16384 Hz. Then, we compute the STFT with a window size of 1022 and a hop length of 256. This leads to a $512 \times 64\, n$ complex spectrogram where $n$ is the duration of the waveform in seconds. To reduce the computational cost of both training and inference we downsample the spectrogram in the frequency dimension by 2 in Stage 1. 

\section{Datasets}  \label{datasets}
Experiments are carried out in two different datasets: \textit{Voxceleb2} \cite{voxceleb}, a dataset of celebrities speaking in a broad range of scenarios; and \textit{Acappella} \cite{montesinos}, a dataset of solo-singing videos. Both datasets are a collection of YouTube recordings which are publicly available. We also consider \textit{Audioset} \cite{gemmeke2017audio} and \textit{MUSDB18} \cite{musdb18}  for sampling extra audio sources that can be added to the singing voice signal as accompaniment.

\textit{Voxceleb2} contains 1 million utterances, most of them of a duration between 4 and 6 seconds, consisting of celebrities covering a wide range of ethnicities, professions and ages. The dataset is formed by in-the-wild videos that include several challenging scenarios, such as: different lightning, side-face views, motion blur and poor image quality. They also span across different scenarios like red carpets, stadiums, public speeches, etc. The dataset provides a test set which contains both, seen-heard and unseen-unheard speakers together. From this test set we selected the unseen-unheard samples and curated two different subsets.  The first one, denoted as \textit{unheard-unseen wild test set} consists of 1,000 samples randomly selected, reflecting the aforementioned challenges. The second one, denoted as \textit{unheard-unseen clean test set}, is a subset of 1,000  samples, from which 500 of them have a high-quality content with the following characteristics: frontal or almost frontal point of view, low background noise and perceptual image quality above the average of the dataset. The samples were selected manually from the whole unseen-unheard test set, trying to include as many different speakers as possible. The target voice is sampled from the subset of 500 high-quality videos in the \textit{clean set}, while the second voice is sampled from the rest of 500 videos. This way we ensure that the video content is good enough to estimate motion features from it and that the ground truth separated audio is reliable,  in the sense that it does not contain background sounds that may produce unfounded performance metrics.

\textit{Acappella} is a 46-hours dataset of a cappella solo singing videos. The videos are divided in four language categories: English, Spanish, Hindi and others. These videos are recorded in a frontal view with no occlussions. It also provides two test sets: the seen-heard test set and the unseen-unheard test set. The former contains videos sampled from the same singers and in the same languages than the training set, whereas the latter contains recordings sampled from new singers in the four language categories plus some new languages. In the test set all the categories are equally represented across languages and gender. This way the algorithms can be tested in challenging  real-world scenarios.

\textit{Audioset} \cite{gemmeke2017audio}  is an in-the-wild large-scale dataset of audio events across more than 600 categories. We gathered the categories related to the human voice and some typical accompaniments. These categories are:
acappella, background music, beatboxing, choir, drum, lullaby, rapping, theremin, whistling and yodelling.

Finally, \textit{MUSDB18}\cite{musdb18} is an audio-only dataset of 150 full-track songs of different styles that includes  original sound sources. 
% to have pop and rock music examples as accompaniment.

\section{Experiments}
 The experiments were carried out in a single RTX 3090 GPU. Each experiment takes around 20 days of training. We used SGD with 0.8 momentum, $10^{-5}$ weight decay and a learning rate of 0.01.
The metrics used for comparing results are Source-to-Distortion Ratio (SDR) and Source-to-Interferences Ratio  (SIR) \cite{sdr}.

\subsection{Audio-visual transformer}\label{multimodal-experiment}

In this experiment, we compare three different versions of the transformer (shown in Fig. \ref{transformer-ablation}) in the \textit{Acappella} dataset. The goal is two-fold: i) Compare the proposed architecture against the state-of-the-art model in singing voice separation \cite{montesinos}; %which also uses face landmarks for extracting the motion features
 and ii) compare the performance of different transformers for the task of  singing voice separation. 

%The first proposal is using the transformer as an auto-encoder, being fed with an audio-visual signal directly. This way we ease the task for the transformer as audio and visual features are temporally aligned by construction. Then, it just have to find relationships through the multi-head self-attention. 
%The second proposal is passing visual features to the encoder and audio features (from the mixture) to the decoder. This way the transformer is, in theory, capable of finding dependencies among the video, among the audio and between both video and audio. Nevertheless, we hypothesise the dependencies between video and audio are local as audio events mostly occur at the same time than visual events. 
%Lastly, we feed the encoder with an audio-visual signal and the decoder with the ground-truth separated audio. Note that this model is slower than previous ones as the model runs recurrently at inference time, going from a time complexity of $\mathcal{O}(n)$ to $\mathcal{O}(n^2)$ where $n$ is the sequence length.

For the sake of comparison, we train our models the same way as in \cite{montesinos}. In short, we create artificial mixtures of 4s of duration by mixing a voice sample from \textit{Acappella} together with an accompaniment sample sourced either from \textit{Audioset} or \textit{MUSDB18}. 
%We restrict to the categories from \textit{Audioset}  related to human voice and music, namely: a cappella, background music, beatboxing, choir, drum, lullaby, rapping, theremin, whistling and yodelling \textcolor{red}{(already mentioned before, remove one of them)}. 
Additionally, a second voice sample from \textit{Acappella} is added 50\% of the times. This results in mixtures that contain one or more voices plus musical accompaniment. 
For this dataset we take 4s audio excerpts and the corresponding 100 video frames from which we extract the face landmarks.% \textcolor{red}{BUT here do we use the same landmarks as in BMVC? the we have to say it} \textcolor{red}{We don't use exactly the same extractor but the graph seems to be the same}. 

Results are shown in Table \ref{transformer-ablation}. From the ablation on the three versions of the transformer, we can conclude that the AV ST-transformer is the best model in terms of both performance and time complexity. Moreover, it can be observed that the three versions of the transformer greatly outperform the results of \cite{montesinos} in terms of SDR, while the AV ST-transformer also outperforms in SIR.

\begin{table}[b]
\centering
\begin{tabular}{l|c|c|c|c}
Model & Y-Net \cite{montesinos}          & AV ST-transformer & V\_A transformer & AV\_A transformer \\ \hline
SDR $\uparrow$   & 6.41                             & \textbf{10.63 $\pm$ 5.86}       &    8.64 $\pm$ 5.89           & 9.98 $\pm$ 5.70  \\
SIR $\uparrow$   & 17.38                            & \textbf{17.67 $\pm$ 7.73}       &    14.70 $\pm$ 7.88          & 16.11 $\pm$  7.42
\end{tabular}
\caption{Ablation study: performance of different ways of feeding a transformer with an audio-visual signal and comparison to Y-Net model \cite{montesinos}.
Evaluated in \textit{Acappella's} unseen-unheard test set.
 Y-Net metrics taken from \textit{Acappella}. In this table $N=4$ (the number of blocks in the transformers) in order to  adapt the number of parameters to the size of \textit{Acappella} dataset.}
\label{transfomer-ablation}
\end{table}
\subsection{Speech separation}
In Section \ref{multimodal-experiment} we found the AV ST-transformer was the best model in terms of time complexity and performance. All the remaining experiments will be carried out with this model. Now we consider the task of AV speech separation and work with \textit{Voxceleb2} dataset. We use 2s audio excerpts which correspond to 50 video frames from which we extracted their face landmarks. In this case, we mix two voice samples from \textit{Voxceleb2} which are normalised with respect to their absolute maximum, so that a  mixture is $x(t)=(s_1(t)+s_2(t))/2$. This normalisation aims to have two voices which are codominant in the mixture and that the waveforms of the mixtures are bounded between -1 and 1. Note that the former characteristic is not always true as \textit{Voxceleb2} samples are sometimes accompanied by other voices or sorts of interference (clapping, music,  etc.). As \textit{Voxceleb2} is a large-scale dataset, and for the sake of comparison, we extended the size of the AV ST-transformer up to 10 encoder blocks and 10 decoder blocks so that the number of parameters of the audio subnetwork is comparable to that of Visual Voice \cite{visualvoice}. We tested the performance of each model in the \textit{unheard-unseen wild} test set and in the \textit{unheard-unseen clean} test set (both described in Section \ref{datasets}). For each test set we randomly made 500 pairs out of the 1,000 samples, ensuring no sample is used more than once.

{\bf Lead Voice Enhancer.}\label{sec:refinement}
The first experiment is an ablation designed to address three main questions. 
i) Compare two different versions of the lead voice enhancer: the audio backbone of Y-Net \cite{montesinos}, which is a 7M-parameter U-Net; and the audio backbone of Visual Voice \cite{visualvoice}, yet another U-Net but with 50M parameters because of a different design. ii) Evaluate the effect of recurrent iterations of the lead voice enhancer. And iii) comparing the results of the 10-block 2-stage AV ST-Transformer against a 18-block 1-stage AV ST-Transformer transformer. The details of this subnetwork are explained in Section \ref{approach:refinement}. We denote our Voice-Visual Transformer as VoViT (the whole network with two stages) and VoViT-s1 the network without the second stage.

The results are shown in Table \ref{tab:refinement}. As we can see, the refinement network improves the results substantially for the 10-block AV ST-Transformer. Successive iterations of the refinement module further reduce the interferences, but the best SDR is achieved with just one iteration. 
%however the intelligibility \textcolor{red}{why intellibility? the SDR considers different aspects .... what measures intellibility is STOI/ESTOI} is reduced as well. 
For the lead voice enhancer, we tried two possible audio-only U-Nets: the U-Net from the Y-Net model \cite{montesinos} and the larger U-Net from Visual Voice \cite{visualvoice}. 
A much larger U-Net does not outperform the smaller one by a large margin. Interestingly, we can observe that adding this module performs better than using the 18-block AV ST-transformer (with around 2 times more parameters). Moreover, this subnetwork can be trained within a day, whereas the 18-block transformer required around a month to train. The reasons behind the lack of improvement of the 18-block transformer are unknown. We observed a  phenomena similar to the so called ``double descent" \cite{double_descent} while training the 10-block transformer, which may be indicative of a complex optimisation process which is worsened in the 18-block case exceeding our computational resources. In the same line, we trained a larger graph convolutional network, comparable in number of parameters to the motion subnetwork of Visual Voice, however the performance dropped.
From this ablation, we can conclude that a 10-block AV ST-transformer with a small U-Net as lead voice enhancer is the best option in terms of performance-latency trade-off. %\cc{which is our primary target (can be removed)}. 
% in the other secion we say:
% Further iterations reduce the interferences (at a lesser extent) but at the cost of introducing more distortion.
% From Grais et al
% We used the proposed combining approach to achieve the advantages of two different source separation estimates using binary and soft masks, where the first
%estimate (binary mask) achieves better separation with less interference than the soft mask, but with distorted sources, while
%the second estimate (soft mask) achieves less separation, but
%with less distortion than the binary mask. Combining the estimates of binary and soft masks using another DNN achieves
%less distortion than each estimate individually and as good separation as the binary mask. 
\begin{table}[ht]
\centering
\begin{tabular}{c|lcc}
                             & & \multicolumn{2}{c}{Wild test set} \\ \hline
&                             & SDR $\uparrow$            & SIR $\uparrow$            \\ \hline
\multirow{5}{*}{\STAB{\rotatebox[origin=c]{90}{10-block}}}
&VoViT-s1          & 9.68            & 15.75           \\
&VoViT (VV in stage 2, $r=1$) & \textbf{10.05}  & 18.30           \\
&VoViT (VV in stage 2, $r=2$)  & 9.77            & \textbf{19.38}  \\
&VoViT (YN in stage 2, $r=1$)  & 10.03           & 18.18           \\
&VoViT (YN in stage 2, $r=2$)  & 9.78            & 19.09           \\ \hline
&18-block VoViT-s1              & 9.27            & 15.53          
\end{tabular}%
\caption{Ablation of different variants of the refinement stage and number of blocks in the transformer of the first stage.  VoViT-s1 stands for the model with just the first stage, $r$ stands for the number of recurrent passes in stage 2. For the stage 2  we considerered both, the Visual Voice's UNet (VV) \cite{visualvoice} and the Y-Net's UNet (YN) \cite{montesinos}.}
\label{tab:refinement}
\end{table}

{\bf Comparison to state-of-the-art methods.}
Next we are going to compare the 10-block AV ST-Transformer  to a state-of-the-art AV speech separation model and audio baselines in the  \textit{Voxceleb2} dataset. The Visual Voice network \cite{visualvoice}  is the current state of the art in speech separation. This network uses 2.55s excerpts, the corresponding 64 video frames cropped around the lips and an image of the whole face of the target speaker. Apart from using lips motion features, it extracts cross-modal face-voice embeddings that complement the motion features and are especially useful when the motion is not reliable or when the appearance of the speakers is different. We also compare the results against Y-Net \cite{montesinos} as it is one of the few papers proposing face landmarks. The original work uses 4s excerpts. As around 160k samples for \textit{Voxceleb2} are shorter, we just adapted the model for working with 2s samples. 

\begin{table}[htb]
\centering
\resizebox{\textwidth}{!}{%
\begin{tabular}{lcc|cc|cc}
\multicolumn{1}{c}{} &
  \multicolumn{2}{c|}{\# parameters} &
  \multicolumn{2}{c|}{Wild Test set} &
  \multicolumn{2}{c}{Clean Test set} \\ \cline{2-7} 
\multicolumn{1}{c}{} &
  \multicolumn{1}{l}{\begin{tabular}[c]{@{}l@{}}Visual \\ Net.\end{tabular}} &
  \multicolumn{1}{l|}{\begin{tabular}[c]{@{}l@{}}Whole \\ Net.\end{tabular}} &
  SDR $\uparrow$ &
  SIR $\uparrow$ &
  SDR $\uparrow$ &
  SIR $\uparrow$ \\ \hline
\begin{tabular}[c]{@{}l@{}}Visual Voice Audio-only\end{tabular} &      -- & 46.14 & 7.7            & 13.6           &      --          &      --          \\
Face Filter \cite{facefilter}                                                           &    --   &    --   & 2.53           &       --         &      --          &        --        \\
%\textit{Afouras et al.}
The conversation \cite{Afouras18}                                                                &   --    &   --    & 8.89           & 14.8           &      --          &      --          \\
\begin{tabular}[c]{@{}l@{}}Visual Voice Motion-only\end{tabular}    & 9.14  & 55.28 & 9.94           & 17             &    --            &       --         \\
\hline 
Y-Net \cite{montesinos}                                                                  & 1.42  & 9.7   & 5.29 $\pm$ 5.06  & 8.45 $\pm$ 6.8           & 5.86 $\pm$ 4.78  & 9.25 $\pm$ 6.44           \\
Visual Voice  \cite{visualvoice}                                                         & 20.38 & 77.75 & 9.92 $\pm$ 3.56  & 16.11 $\pm$ 4.8 & 10.18 $\pm$ 3.36     & 16.49 $\pm$ 4.5           \\
%VoViT-s1                                                  & 1.42  & 50.1  & 9.68$\pm$3.31  & 15.75$\pm$4.53 & 9.85$\pm$2.73  & 16.14$\pm$3.84         \\
VoViT                                                  & 1.42  & 58.2  & \textbf{10.03 $\pm$ 3.35} & \textbf{18.18 $\pm$ 4.72} & \textbf{10.25 $\pm$ 2.61} & \textbf{18.65 $\pm $3.8} \\
%VoViT fp16                                               & 1.42  & 58.2  & \textbf{10.03$\pm$3.35} & \textbf{18.18$\pm$4.72} & \textbf{10.25$\pm$2.61} & \textbf{18.65$\pm$3.8} 
\end{tabular}%
}
\caption{Evaluation on \textit{Voxceleb2} unheard-unseen test sets (mean $\pm$ standard deviation). VoViT stands for our model with the 10-block AV ST-Transformer  with the Y-Net's UNet backbone as the lead voice enhancer. Number of parameters in millions.
Results in the first block are taken from the original papers.}  %and that is why we only report certain columns. }
%In the second block we report the mean and the standard deviation values.}
%\sout{\fix{}{The difference between the mean SDR of VoViT and Visual Voice is not significant (p-value$$>$$0.05).} }} %voy a ponerlo en el texto mejor
\label{tab:results}
\end{table}
\begin{figure}[]
  \centering
  \includegraphics[width=0.8\textwidth]{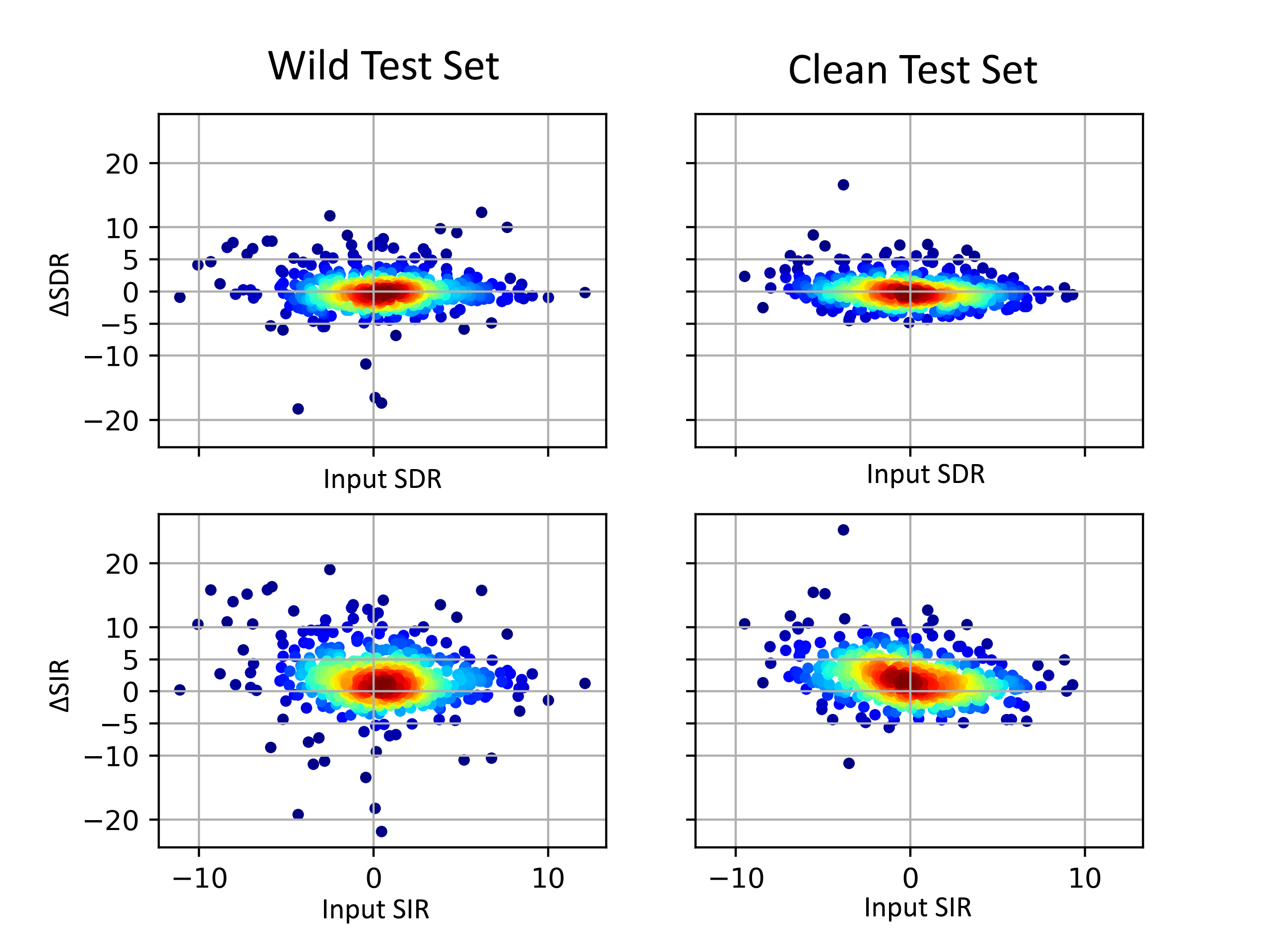}
  \caption{Scatter plot showing the difference in SDR and SIR, $\Delta SDR$ and $\Delta SIR$, as functions of the SDR and SIR of the input mixture in the unseen-unheard wild and clean test sets. The difference is: 
  $\Delta SDR = SDR(\text{VoViT})-SDR(\text{Visual Voice})$ so a positive value means VoViT outperforms Visual Voice.}
  \label{scatter_plots}
\end{figure}

Numerical results are shown in Table \ref{tab:results}. The 10-block VoViT outperforms all the previous AV speech separation models. Compared to Visual Voice, it achieves a much better SIR and slightly better SDR, both for the wild and clean test sets. In particular, for the clean test set, when the motion cues are more reliable, our model has a much lower standard deviation. Some aspects need to be taken into account:
\begin{comment}
\begin{itemize}
    \item The face landmark extractor has been trained with higher quality videos than the ones in \textit{Voxceleb2}. On the contrary, the Visual Voice video network has been trained specifically for \textit{Voxceleb2}.
    \item Our visual subnetwork, the graph CNN, has 10 times less parameters than its counterpart in Visual Voice. 
    \item Apart from motion cues, Visual Voice takes also into account speaker appearance features which are correlated with voice features, and which can be crucial in poor quality videos where lip motion is unreliable.
\end{itemize}
\end{comment}

- The face landmark extractor has been trained with higher quality videos than the ones in \textit{Voxceleb2}. On the contrary, the Visual Voice video network has been trained specifically for \textit{Voxceleb2}.

- Our visual subnetwork, the graph CNN, has 10 times less parameters than its counterpart in Visual Voice.

- Apart from motion cues, Visual Voice takes also into account speaker appearance features which are correlated with voice features, and which can be crucial in poor quality videos where lip motion is unreliable.

Fig. \ref{scatter_plots} shows SDR and SIR differences between VoViT and Visual Voice in two different test sets: the \textit{wild} and the \textit{clean} set. Each plot is a scatter plot where each point corresponds to a 2s long mixture. 
%The plots depict the SDR/SIR difference between both methods as a function of the SDR/SIR of the mixture. A positive value in the increment of SDR/SIR means that our result is better than the one from Visual Voice.
As it can be observed, our method especially outperforms Visual Voice in SIR while in SDR both methods have a comparable performance. 
In order to assess the significance of the results of Table \ref{tab:results}, we calculated the $p$-values with respect to the Visual Voice results. Only the improvement on SIR is significant ($p < 0.05$). While the improvement from stage 1 to 2 (Table \ref{tab:refinement}) is significant both in SDR and SIR.
In the \textit{wild test set} there are a few samples where our model performs worse than Visual Voice. Those correspond to samples where the audio and video are extremely unsynchronised or samples where the lip motion is mispredicted and the network separates the other speaker. In those cases, the Visual Voice model might be able to alleviate the situation either by relying on the appearance features to guide the separation or by using the motion information present in the raw video despite its poor quality
(e.g. blur, compression artefacts, lack of sharpness). There are no such cases in the \textit{clean set}, as those type of samples were filtered out. Audio-visual files with the top $K$ worst performing examples and demos for both, Visual Voice and VoViT models, are provided in the supplementary material.

\begin{comment}
\begin{table}[ht]
\centering
\resizebox{\textwidth}{!}{%
\begin{tabular}{lc|cc|c}
 & Preprocessing & \multicolumn{2}{c|}{Inference} & \begin{tabular}[c]{@{}c@{}}Preprocessing\\  + Inference\end{tabular} \\ \cline{3-4}
                                                       &       & Graph Network & Whole model &       \\ \hline
\multicolumn{1}{l|}{VoViT-s1}      & 17.95 & 4.50          & 52.21       & 82.18 \\
\multicolumn{1}{l|}{VoViT}       & 17.95 & 4.55          & 57.45       & 93.31 \\
\multicolumn{1}{l|}{VoViT-s1 fp16} & 10.94 & 2.88          & 30.47       & 52.43 \\
\multicolumn{1}{l|}{VoViT fp16} & 10.94 & 2.86          & 34.18       & 46.14
\end{tabular}%
}
\caption{Latency estimation for the different variants of VoViT. Average of 10 runs, batch size 100. Device: Nvidia RTX 3090. GPU utilization \textgreater{}98\%, memory on demand. Two forward passed done to warm up. 
Timing corresponds to ms to process 10s of audio. \textcolor{red}{remove this table}}
\label{tab:latency}
\end{table}
\end{comment}
\subsection{Singing voice separation}
In this last experiment we consider the task of singing voice. We are interested in exploring how transferable models trained for speech separation are to the case of singing voice. Since speech models were trained with two voices and no extra sounds and in \textit{Voxceleb2}, which contains mainly English, we restricted to similar types of mixtures in singing voice. In particular, we create mixtures of two singers in English from the unseen-unheard test set of \textit{Acappella}, with no accompaniment.  Table \ref{tab:singing_voice} compares the results of models trained directly with samples of singing voice (top block of results in Table \ref{tab:singing_voice}) versus models trained with speech samples (bottom block). In the case of singing voice we used our model with just the first stage and a 4-block AV ST-transformer. 
We observe that dedicated models for singing voice perform largely better than models trained for speech. This may be explained to particular differences between a speaking and a singing voice. For example, vowels are much more sustained in singing voice, there is much less coarticulation of consonants with surrounding vowels and vibrato is not present in speech. Moreover, singing voice contains varying pitches covering a wider frequency range.
%\textcolor{red}{somme comments about table \ref{tab:singing_voice}, a model trained for speech does not generalise good enough to singing voice, better to train specifically for singing voice, hypothesis: differences between speaking and singing. Vowels are much more sustained in singing voice, vibrato not present in speech (but do we have vibratos in acappela?) Singing has varying pitches and rhythm. Extracted from commented link below: "In general, singing
%occurs at higher frequencies levels than speech and covers a much wider frequency range. In singing, there is much less coarticulation of consonants with surrounding vowels. Because singing has a wider frequency and intensity range than speech, features vowels which are long in duration, and a larger overall vocal tract, one could say that singing is indeed “more resonant” than speech."}
% http://music.utsa.edu/pdfs/61_SpeakingvsSinging.pdf

\begin{table}[h]
\centering
\begin{tabular}{l|c|c}
Model & SDR  $\uparrow$  & SIR $\uparrow$  \\ 
\hline
Y-Net \cite{montesinos} &  11.08 $\pm$ 7.51 & 17.18 $\pm$ 9.68 \\
VoViT-s1 (4 blocks) & \textbf{14.85 $\pm$ 7.87}  & \textbf{21.06 $\pm$ 9.69}  \\
\hline                           
VoViT-s1 & 3.89 $\pm$ 9.28   &  5.89 $\pm$ 11.15 \\
VoViT &  4.04 $\pm$ 10.30 &  \textbf{7.21 $\pm$ 13.26} \\
Visual Voice \cite{visualvoice} & \textbf{4.52 $\pm$ 8.64} & 7.03 $\pm$ 7.11
\end{tabular}
\caption{Singing voice separation. Mixtures of two singers with no additional accompaniment from the test set unseen-unheard (only samples in English) of \textit{Acappella}. Results in top block: models trained directly with samples of singing voice; bottom block: models trained with speech samples.}
\label{tab:singing_voice}
\end{table}

\section{Conclusions \& Future work}
In this work we present a lightweight audio-visual source separation method which can process 10s of recordings in less than 0.1s in an end-to-end GPU powered manner. Besides, the method shows competitive results to the state-of-the-art in reducing distortions while clearly outperforming in reducing interferences. 
We show that face landmarks are computationally cheaper alternatives to raw video and help to deal with large-scale datasets. 
For the first time, we evaluate AV speech separation systems in singing voice, showing empirically that the characteristics of the singing voice differ substantially from the ones of speech. 

As future work we would like to explore lighter and faster embedding generators for the transformer and different optimisations in its architecture which leads to a fast and powerful system.   

\vspace{0.2cm}
\noindent {
 \textbf{Acknowledgments.}
We acknowledge support by MICINN/FEDER UE project PID2021-127643NB-I00; H2020-MSCA-RISE-2017 project  777826 NoMADS. 
 %V.\ S.\ K.\ has received support through “la Caixa” Foundation (ID 100010434), fellowship code: LCF/BQ/DI18/11660064 
%and the Marie SkłodowskaCurie grant agreement No.\ 713673. 

\noindent J.F.M. acknowledges support by FPI scholarship PRE2018-083920.%V.S.K has also received funding from the European Union’s Horizon 2020 research and innovation programme under the Marie SkłodowskaCurie grant agreement No. 713673. 
We acknowledge NVIDIA Corporation for the donation of GPUs used for the experiments.}
\clearpage
\bibliographystyle{splncs04}
\bibliography{egbib}
\end{document}